\documentclass[english,prstab,twocolumn,superscriptaddress,longbibliography]{revtex4-1}

\usepackage{xcolor}
\usepackage{graphicx}
\usepackage{url}
\usepackage{amsmath}
\usepackage{amsfonts}
\usepackage{color}
\usepackage[utf8]{inputenc}
\usepackage{babel}

\newcommand{\bs}[1]{\boldsymbol{#1}}

\begin{document}
 
\title{Analysis of beam position monitor requirements with Bayesian
  Gaussian regression}
 
\author{Yongjun Li} \thanks{Email: yli@bnl.gov}
\affiliation{Brookhaven National Laboratory, Upton, New York 11973, USA}

\author{Yue Hao}
\affiliation{FRIB/NSCL, Michigan State University, East Lansing, Michigan 48864, USA}

\author{Weixing Cheng}
\affiliation{Brookhaven National Laboratory, Upton, New York 11973, USA}
\affiliation{Argonne National Laboratory, Argonne, Illinois 60439, USA}

\author{Robert Rainer}
\affiliation{Brookhaven National Laboratory, Upton, New York 11973, USA}

 
\begin{abstract}
  With a Bayesian Gaussian regression approach, a systematic method for
  analyzing a storage ring's beam position monitor (BPM) system
  requirements has been developed.  The ultimate performance of a
  ring-based accelerator, based on brightness or luminosity, is determined
  not only by global parameters, but also by local beam properties at some
  particular points of interest (POI).  BPMs used for monitoring the beam
  properties, however, cannot be located at these points.  Therefore, the
  underlying and fundamental purpose of a BPM system is to predict whether
  the beam properties at POIs reach their desired values.  The prediction
  process can be viewed as a regression problem with BPM readings as the
  training data, but containing random noise.  A Bayesian Gaussian
  regression approach can determine the probability distribution of the
  predictive errors, which can be used to conversely analyze the BPM
  system requirements.  This approach is demonstrated by using
  turn-by-turn data to reconstruct a linear optics model, and predict the
  brightness degradation for a ring-based light source.  The quality of
  BPMs was found to be more important than their quantity in mitigating
  predictive errors.
\end{abstract}
 
\maketitle

\section{\label{sect:intro}introduction}
  The ultimate performance of a ring-based accelerator is determined not
  only by certain critical global parameters, such as beam emittance, but
  also by local properties of the beam at particular points of interest
  (POI).  The capability of diagnosing and controlling local beam
  parameters at POIs, such as beam size and divergence, is crucial for a
  machine to achieve its design performance.  Examples of POIs in a
  dedicated synchrotron light source ring include the undulator locations,
  from where high brightness X-rays are generated.  In a collider, POIs
  are reserved for detectors in which the beam-beam luminosity is
  observed.  However, beam diagnostics elements, such as beam position
  monitors (BPM) are generally placed outside of the POIs as the POIs are
  already occupied.  An intuitive, but quantitatively unproven belief, is
  that the desired beam properties at the POIs can be achieved once the
  beam properties are well-controlled at the location of the BPMs.

  Using observational data at BPMs to indirectly predict the beam
  properties at POIs can be viewed as a regression problem and can be
  treated as a supervised learning process: BPM readings at given
  locations are used as a training dataset.  Then a ring optics model with
  a set of quadrupole excitations as its arguments is selected as the
  hypothesis.  From the dataset, an optics model needs to be generalized
  first.  Based on the model, the unknown beam properties at POIs can be
  predicted.  However, there exists some systematic error and random
  uncertainty in the BPMs' readings, and the quantity of BPMs (the
  dimension of the training dataset) is limited.  Therefore, the
  parameters in the reconstructed optics model have inherent
  uncertainties, as do the final beam property predictions at the POIs.
  The precision and accuracy of the predictions at the POIs depend on the
  quantity of BPMs, their physical distribution pattern around the ring,
  and their calibration, resolution, etc.  When a BPM system is designed
  for a storage ring, however, it is more important to consider the
  inverse problem: i.e. How are the BPM system technical requirements
  determined in order to observe whether the ring achieves its desired
  performance?  In this paper, we developed an approach to address this
  question with Bayesian Gaussian regression.

  In statistics, a Bayesian Gaussian
  regression~\cite{Rasmussen,Bishop:2006} is a Bayesian approach to
  multivariate regression, i.e. regression where the predicted outcome is
  a vector of correlated random variables rather than a single scalar
  random variable.  Every finite collection of the data has a normal
  distribution.  The distribution of generalized arguments of the
  hypothesis is the joint distribution of all those random variables.
  Based on the hypothesis, a prediction can be made for any unknown
  dataset within a continuous domain.  In our case, multiple BPMs'
  readings are normally distributed around their real values.  The
  standard deviations of the Gaussian distributions are BPM's resolutions.
  A vector composed of quadrupoles' mis-settings is the argument to be
  generalized.  The prediction at the POIs is the function of this vector.
  The continuous domain is the longitudinal coordinate $s$ along a storage
  ring.

  To further explain this approach, the remaining sections are outlined as
  follows: Sect.~\ref{sect:performance} introduces the relation between
  machine performance and beam diagnostics system capabilities.
  Sect.~\ref{sect:GPmodel} explains the procedure of applying the Bayesian
  Gaussian regression in the ring optics model reconstruction, and the
  prediction of local optics properties at POIs.  In
  Sect.~\ref{sect:nsls2}, the National Synchrotron Light Source II
  (NSLS-II) storage ring and its BPM system are used to illustrate the
  application of this approach.  Some discussions and a brief summary is
  given in Sect.~\ref{sect:summary}.

\section{machine performance and beam diagnostics capability\label{sect:performance}}

  As mentioned previously, ultimate performance of a ring-based
  accelerator relies heavily on local beam properties at particular POIs.
  Consider a dedicated light source ring.  Its ultimate performance is
  measured by the brightness of the X-rays generated by undulators.  The
  brightness of undulator emission is determined by the transverse size of
  both the electron and photon beam and their angular divergence at their
  source points~\cite{Lindberg,Walker,Chubar,Hidas}.  Therefore, the
  undulator brightness performance $\mathcal{B}$ depends on the ring's
  global emittance and the local transverse optics parameters,
  \begin{eqnarray}
  \mathcal{B} & \propto &
  \frac{1}{\Sigma_x\Sigma_x^{\prime}\Sigma_y\Sigma_y^{\prime}}
  \nonumber\\ \Sigma_{x,y} & = &
  \sqrt{\epsilon_{x,y}\beta_{x,y}+\eta_{x,y}^2\sigma_{\delta}^2+\sigma^2_{ph}}
  \nonumber\\ \Sigma_{x,y}^{\prime} & = &
  \sqrt{\epsilon_{x,y}\gamma_{x,y}+\eta_{x,y}^{\prime2}
    \sigma_{\delta}^2+\sigma^{\prime2}_{ph}}. \label{eq:brightness}
  \end{eqnarray}
  Here $\epsilon_{x,y}$ are the electron beam emittances, which represent
  the equilibrium between the quantum excitation and the radiation damping
  around the whole ring.  $\beta,\gamma$ are the Twiss
  parameters~\cite{Courant:1997rq}, $\eta,\eta^{\prime}$ are the
  dispersion and its derivative at the undulators' locations,
  $\sigma_{\delta}$ is the electron beam energy spread
  $\sigma_{ph}=\frac{\sqrt{\lambda L_u}}{2\pi}$ and
  $\sigma^{\prime}_{ph}=\frac{1}{2}\sqrt{\frac{\lambda}{L_u}}$ are the
  X-ray beam diffraction ``waist size'' and its natural angular
  divergence, respectively.  The X-ray wavelength $\lambda$, is determined
  based on the requirements of the beam-line experiments, and $L_u$ is the
  undulator periodic length.  The emittance was found to be nearly
  constant with small $\beta$-beat (see Sect.~\ref{sect:nsls2}).
  Therefore, monitoring and controlling the local POI's Twiss parameters
  is crucial.

  The final goal of beam diagnostics is to provide sufficient, accurate
  observations to reconstruct an online accelerator model.  Modern BPM
  electronics can provide the beam turn-by-turn (TbT) data, which is
  widely used for the beam optics characterization and the model
  reconstruction.  Based on the model, we can predict the beam properties
  not only at the locations of monitors themselves, but more importantly
  at the POIs.  The capability of indirect prediction of the Twiss
  parameters at POIs eventually defines the BPM system requirements on TbT
  data acquisition.  Based on Eq.~\eqref{eq:brightness}, how precisely one
  can predict the bias and the uncertainty of Twiss parameters $\beta$ and
  $\eta$ at locations of undulators is the key problem in designing a BPM
  system.  Therefore, to specify the technical requirements of a BPM
  system, the following questions need to be addressed: in order to make
  an accurate and precise prediction of beam properties at POIs, how many
  BPMs are needed? How should the BPMs be allocated throughout the
  accelerator ring, and how precise should the BPM TbT reading be?

  In the following section a method of reconstructing the linear optics
  model, and determining the brightness performance for a ring-based light
  source will be discussed.  For a collider ring, its luminosity is
  determined only by the beam sizes at the interaction
  points~\cite{Herr:2003em}.  Gaussian regression analysis can therefore
  be applied to predict its $\beta^*$ and luminosity as well.

  \section{\label{sect:GPmodel}Gaussian regression for model reconstruction and prediction}

\begin{figure}[!ht]
  \centering \includegraphics[width=1.\columnwidth]{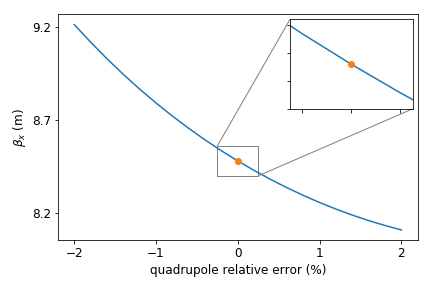}
  \caption{\label{fig:linear} $\beta_x$ dependency on the excitation error
    of a quadrupole observed by a BPM at the NSLS-II ring. The dependency
    is nonlinear.  However, when the quadrupole error is confined to a
    small range $[-0.25\%,0.25\%]$, it can be approximated as a linear
    dependence as shown in the zoomed-in window.  At modern storage rings,
    such as NSLS-II, quadrupole excitation errors due to a power supply's
    mis-calibration and/or magnetic hysteresis are much less than 0.25\%.}
  \end{figure}

  When circulating beam in a storage ring is disturbed, a BPM system can
  provide its TbT data at multiple longitudinal locations.  TbT data of
  the BPMs can be represented as an optics model plus some random reading
  errors,
  \begin{equation}\label{eq:tbt}
  x(s)_i =
  A(i)\sqrt{\beta(s)}\cos\left[i\cdot2\pi\nu+\phi(s)\right]+\varepsilon(s)_i,
  \end{equation}
  here $i$ is the index of turns, $A(i)$ is a variable dependent on turn
  number, $\beta(s)$ is the envelope function of Twiss parameters at $s$
  location, $\nu$ is the betatron tune, $\phi$ is the betatron phase, and
  $\varepsilon(s)_i$ is the BPM reading noise~\cite{Calaga,Langner,Cohen},
  which generally has a normal distribution.  Based on the accelerator
  optics model defined in Eq.~\eqref{eq:tbt}, we can extract a set of
  optics Twiss parameters at all BPM
  locations~\cite{Castro:1996,Irwin:1999,Huang:2005,Tomas:2017}.
  Recently, Ref.~\cite{Hao:2019lmn} proposed using a Bayesian approach to
  infer the mean (aka expectation) and uncertainty of Twiss parameters at
  BPMs simultaneously.  The mean values of $\beta$ represent the most
  likely optics pattern.  The random BPM reading error and the
  simplification of the optics model can result in some uncertainties,
  $\varepsilon_{\beta}$, in the inference process,
  \begin{equation}\label{eq:beta}
  \beta = \beta(s,\bs{K})+\varepsilon_{\beta}(s),
  \end{equation}
  here $\bs{K}$ is a vector composed of all normalized quadrupole focusing
  strengths, and $\varepsilon_{\beta}$ is the inference uncertainty.
  Unless otherwise stated, bold symbols, such as ``$\bs{X}$'', are used to
  denote vectors and matrices throughout this paper.  In accelerator
  physics, the deviation from the design model $\beta_0$ is often referred
  to as the $\beta$-beat.  From the point of view of model reconstruction,
  the $\beta$-beat is due to quadrupole excitation errors and can be
  determined by
  \begin{equation}\label{eq:beta-beat}
  \Delta\beta=\beta(s,\bs{K}_0+\Delta\bs{K})-\beta_0(s,\bs{K}_0)
  \approx\bs{M}\Delta\bs{K},
  \end{equation}
  where $\bs{K}_0$ represents the quadrupoles' nominal setting and
  $\beta_0$ is the nominal envelope function along $s$.  $\bs{M}$ is the
  response matrix composed of elements
  $M_{i,j}=\frac{\partial\beta_{s_i}}{\partial K_j}$ observed by the BPMs.
  The dependency of $\beta$ on $\bs{K}$ is not linear in a complete optics
  model.  However, when quadrupole errors are small enough, the dependence
  can be approximated as a linear relation as illustrated in
  Fig.~\ref{fig:linear}.  The approximation holds for most operational
  storage rings, and other diffraction limited light sources under design
  or construction.  A linear approximation allows us to use the linear
  regression approach for this process.  Eq.~\eqref{eq:beta} or
  \eqref{eq:beta-beat} is a hypothesis with the unknown arguments $\bs{K}$
  or $\Delta\bs{K}$, which need to be generalized from BPM measurement
  data.
  
  Given a set of measured optics parameters $\beta$s at multiple locations
  $s$ from BPM TbT data, the posterior probability of the quadrupole error
  distribution $p(\Delta\bs{K}|\bs{\beta})$ can be given according to
  Bayes theorem~\cite{Li:2019kch},
   \begin{eqnarray}
    p(\Delta\bs{K}|\bs{\beta}) & = &
    \frac{p(\bs{\beta}|\Delta\bs{K})
    p(\Delta\bs{K})}{p(\bs{\beta})} \nonumber \\ &
    \propto &
    p(\bs{\beta}|\Delta\bs{K})p(\Delta\bs{K}). \label{eq:bayes}
   \end{eqnarray}
  Here $p(\bs{\beta}|\Delta\bs{K})$ is referred to as the likelihood function,
  \begin{eqnarray}
   \mathcal{N}(\bs{\beta}|\bar{\bs{\beta}},\sigma_{\beta}^2) & = &
   \frac{1}{\sqrt{2\pi}\sigma_{\beta}}
   \exp\left[-\frac{(\bs{\beta}-\bar{\bs{\beta}})^2}{2\sigma_{\beta}^2}\right]
   \nonumber \\ 
   & \approx & \frac{1}{\sqrt{2\pi}\sigma_{\beta}}
   \exp\left[-\frac{(\Delta\bs{\beta}-\bs{M}\Delta
      \bs{K})^2}{2\sigma_{\beta}^2}\right]. \label{eq:likelihood}
  \end{eqnarray}
  Here $\bar{\beta}=\mathbb{E}(\beta)$ and $\sigma^2_{\beta}$ are the
  expectation value and the variance of the normal distribution of
  measured $\beta$s.  Once the expectation value of the optics measurement
  is extracted from the TbT data, a prior quadrupole excitation error
  distribution $p(\Delta\bs{K})$ can be determined by comparing them
  against the design optics model,
  \begin{eqnarray}\label{eq:prior}
    p(\Delta\bs{K}) &=&\mathcal{N}(\Delta\bs{K}|0,\sigma_{\Delta
      K}^2) \nonumber \\ & = & \frac{1}{\sqrt{2\pi}\sigma_{\Delta K}}
    \exp\left[-\frac{\Delta\bs{K}^2}{2\sigma^2_{\Delta K}}\right],
  \end{eqnarray}
  in which the variance $\sigma^2_{\Delta K}$ of the prior distribution
  $p(\Delta\bs{K})$ is linearly proportional to the mean value of the
  measured $\beta$-beat,
  \begin{equation}\label{eq:prior2}
    \sigma_{\Delta K} \sim \kappa|\Delta\beta| =
    \kappa|\bar{\beta}-\beta_0|.
  \end{equation}
  Here ``$\sim$'' in Eq.~\eqref{eq:prior2} describes a statistically
  proportional relationship between $\beta$-beats (in the unit of ``m'')
  and quadrupole strength error $\Delta K$ (in units of $m^{-2}$).  The
  coefficient $\kappa$ can be computed based on the optics model either
  analytically or numerically before carrying out any measurements.  In the
  NSLS-II ring, $\kappa \approx 1.6\times10^{-3} m^{-3}$, i.e. a $0.25 m$
  $\beta$-beat ($\frac{\Delta\beta}{\beta}\approx1\%$) corresponds to a
  distribution of quadrupole errors with the standard deviation
  $4\times10^{-4} m^{-2}$ ($\frac{\Delta K}{K}\approx0.12\%$) as shown in
  Fig. 1 in Ref. ~\cite{Li:2019kch}.  Qualitatively, the relative
  $\beta$-beat and quadrupole error, i.e. $\frac{\Delta\beta}{\beta}$ and
  $\frac{\Delta K}{K}$ are often used in accelerator literature.
  Here the absolute $\Delta\beta$ and $\Delta K$ are used simply because
  they were adapted to our quantitative implementation.

  Both the likelihood function and the prior distribution are generally
  normally distributed.  Therefore, the posterior distribution is a normal
  distribution by summing over the arguments of the exponentials in
  Eq.~\eqref{eq:likelihood} and \eqref{eq:prior},
  \begin{equation}\label{eq:posterior1}
  (\Delta\bs{\beta}-\bs{M}\Delta\bs{K})^T\bs{S}_{\beta}^{-1}
  (\Delta\bs{\beta}-\bs{M}\Delta\bs{K})+
  \Delta\bs{K}^T\bs{S}_{K}^{-1}\Delta\bs{K}.
  \end{equation}
  Here
  \begin{equation}\label{eq:S}
    \bs{S}_{\beta}^{-1} = \frac{1}{\sigma_{\beta}^2}\bs{I}, \;\;
    \bs{S}_{K}^{-1} = \frac{1}{\sigma_{\Delta K}^2}\bs{I}.
  \end{equation}
  The identity matrix $\bs{I}$ is used in Eq.~\eqref{eq:S} because all
  BPMs' resolutions are assumed to have the same values $\sigma_{\beta}$.
  In reality, however, $\bs{S}_{\beta}^{-1}$ needs to be replaced with a
  diagonal matrix with different elements if the BPMs' resolutions are
  different.  The quadrupoles' error distribution matrix $\bs{S}_{K}^{-1}$
  needs to be processed in the same way if necessary.  The mean value of
  the posterior, corresponding to the most likely quadrupole error
  distribution, can be used to implement the linear optics correction as
  explained in Ref.~\cite{Li:2019kch},
  \begin{equation}\label{eq:posteriorMean}
    \bs{m} =
    \sigma^{-2}_{\beta}\bs{A}^{-1}
    \bs{M}^T\Delta\bs{\bar{\beta}},
  \end{equation}
  where
  $\bs{A}=\left[\sigma^{-2}_{\beta}\bs{M}^T\bs{M}+\sigma^{-2}_{\Delta
      K}\bs{I}\right]$. Adding an extra term $\sigma^{-2}_{\Delta
    K}\bs{I}$ to prevent overfitting is known as the regularization
  technique.  The posterior variance represents the uncertainty of
  quadrupole errors.
  \begin{equation}\label{eq:posteriorVariance}
  \bs{\Sigma_K^2}=\bs{A}^{-1}.
  \end{equation}
  Given $\beta$-beats observed at $s$, the posterior generalizes an optics
  model, in which the quadrupoles errors are normally distributed,
  \begin{equation}\label{eq:posterior2}
    p(\Delta\bs{K}|\Delta\bs{\beta},s)=\mathcal{N}(\Delta\bs{K}|\bs{m},\bs{\Sigma_K^2}),
  \end{equation}
  with the mean value and the variance given by
  Eq.~\eqref{eq:posteriorMean} and \eqref{eq:posteriorVariance}
  respectively.

  Thus far, the optics are measured at the locations of the BPMs, and the
  corresponding quadrupole error distributions are generalized based on
  the measurements.  To confirm the machine brightness performance, we
  need to predict the beam properties at POIs.  To do so, the output of
  all possible posterior quadrupole error distributions must be averaged,
  \begin{eqnarray}
  p(\Delta\bs{\beta}_*|\bs{s}_*,\Delta\bs{\beta},\bs{s}) & = & \int
  p(\Delta\bs{\beta}_*|\bs{s}_*,\Delta\bs{K})
  p(\Delta\bs{K}|\Delta\bs{\beta},\bs{s}) \mathrm{d}\bs{K} \nonumber \\ & =
  & \mathcal{N}(\bs{m}_*,\bs{\Sigma}_*^2). \label{eq:prediction}
  \end{eqnarray}
  Here $\Delta\bs{\beta}_*$ is the predicted result at POIs' locations
  $\bs{s}_*$ given the measured $\Delta\bs{\beta}$ at $\bs{s}$.  The mean
  values and the variances of the predicted distributions at POIs are
  \begin{eqnarray}
  \bs{m}_* & = & \sigma^{-2}_{\beta}\bs{M}_*\bs{A}^{-1}\bs{M}^T\Delta\bs{\bar{\beta}} \nonumber \\
  \bs{\Sigma}_*^2 & = &\bs{M}_*\bs{A}^{-1}\bs{M}^T_*, \label{eq:predictionMeanVar}
  \end{eqnarray}
  $\bs{M}_*$ is the Jacobian matrix of the optics response to quadrupole
  errors observed at POIs.  The difference between the mean value
  $\bs{m}_x$ and the real $\beta$ at a POI is referred to as the
  predicted bias.  By substituting the bias and uncertainty back into
  Eq.~\eqref{eq:brightness}, we can estimate how accurate the brightness
  could be measured for given BPMs' resolutions.  Based on the desired
  brightness resolution, we can determine the needed quantity and
  resolution of BPMs.

\section{\label{sect:nsls2} Application to NSLS-II ring}
  In this section, we use the NSLS-II ring and its BPM system TbT data
  acquisition functionality to demonstrate the application of this
  approach.  NSLS-II is a $3^{rd}$ generation dedicated light source.  All
  undulator source points (POIs) are located at non-dispersive straights.
  Typical photon energy from undulators is around 10 $keV$, with
  corresponding wavelengths around 0.124 $nm$.  The undulators' period
  length is 20 $mm$.  The horizontal beam emittance is 0.9 $nm\cdot rad$
  including the contribution from 3 damping wigglers. The emittance
  coupling ratio can be controlled to less than 1\%.  At its 15 short
  straight centers, the Twiss parameters are designed to be as low as
  $\beta_{x,y}=1.80,\,1.20\,m$, and $\alpha_{x,y}=0$ to generate the
  desired high brightness x-ray beam from the undulators.

\begin{figure}[!ht]
  \centering \includegraphics[width=1.\columnwidth]{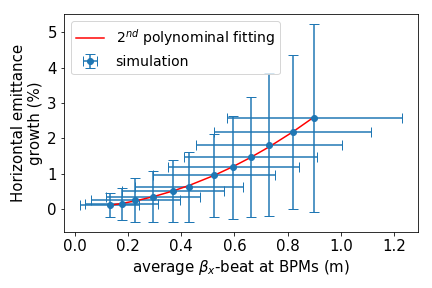}
  \caption{\label{fig:emit_vs_bb} Beam horizontal emittance growth with
    the average $\beta_x$-beat for the NSLS-II ring. If the global
    $\beta_x$-beat can be controlled within 1\%($\sim 0.15m$), the
    emittance growth is negligible.}
  \end{figure}

  The horizontal emittance growth with an optics distortion was studied by
  carrying out a lattice simulation.  With $\beta$-beat at a few percent,
  the corresponding $\alpha-$ and $\gamma$-distortions were generated by
  adding some normally distributed quadrupole errors based on
  Eq.~\eqref{eq:prior} and \eqref{eq:prior2}.  The horizontal emittance
  was found to grow slightly with the average $\beta$-beats as
  illustrated in Fig.~\ref{fig:emit_vs_bb}.  When there is about a 1\%
  horizontal $\beta$-beat ($\sim 0.14 m$), the emittance increases by only
  about 0.1\%, which is negligible.  Therefore, in the following
  calculation, the emittance was represented as a constant.

\begin{figure}[!ht]
  \centering \includegraphics[width=1.\columnwidth]{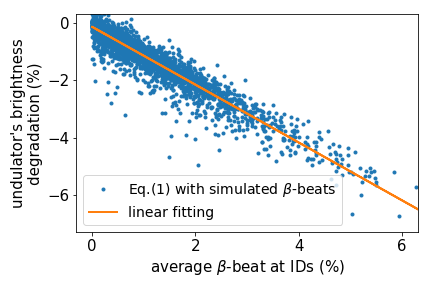}
  \caption{\label{fig:brg_vs_bb} Brightness degradation of an undulator at
    a low-$\beta$ straight due to the average $\beta$-beats in the
    horizontal and vertical planes. Each dot represents a set of simulated
    optics distortions. The brightness degradation has an approximate
    linear dependence on $\beta$-beat.}
  \end{figure}

  Degradation of an undulatorís brightness is determined by its local
  optics distortion which can be evaluated with Eq.~\eqref{eq:brightness}.
  Multi-pairs of simulated $\beta-\alpha$ were incorporated into the
  previously specified undulator parameters to observe the dependence of
  the X-ray brightness on the $\beta$-beat (see Fig.~\ref{fig:brg_vs_bb}).
  A change of approximately 1\% of the $\beta_{x,y}$ in the transverse
  plane can degrade the brightness by about 1\%.  In other words, in order
  to resolve a 1\% brightness degradation, the predictive errors of the
  ring optics (including the bias and uncertainty) at the locations of
  undulators should be less than 1\%.  Because multiple undulators are
  installed around the ring, the predicted performance needs to be
  evaluated at all POIs simultaneously.

  There exist two types of errors in Eq.~\eqref{eq:tbt} which can
  introduce uncertainties in characterizing the optics parameters at BPMs.
  First, due to radiation damping, chromatic decoherence and nonlinearity,
  a disturbed bunched-beam trajectory is not a pure linear undamping
  betatron oscillation~\cite{Meller:1987ug}.  A reduced model (for
  example, assuming $A$ is a constant), will introduce systematic
  errors~\cite{Malina,Carla,Franchi,Langner}.  The second error source is
  the BPM TbT resolution limit, which results in random noise.  At
  NSLS-II, the BPM TbT resolution at low beam current ($<2\; mA$) is
  inferred as $\sim10-15\;\mu m$.  When a $2^{nd}$ order polynomial
  function is used to represent the turn-dependent amplitude $A(i)$, the
  inferred $\beta$ function resolution at BPMs can be reached as low as
  0.5\%~\cite{Hao:2019lmn}.

  First we studied the dependence of predictive errors on the quantity of
  BPMs.  A comprehensive simulation was set up to compare the Gaussian
  regression predictive errors with the real errors.  A linear optics
  simulation code was used to simulate the distorted optics due to a set
  of quadrupole errors.  The $\beta$-beats observed at the BPMs were marked
  as the ``real'' values.  On top of the real values, 0.5\% random errors
  were added to simulate one-time measurement uncertainty seen by the
  BPMs.  A posterior distribution Eq.~\eqref{eq:posteriorMean} and
  \eqref{eq:posteriorVariance} of the quadrupole errors was obtained by
  reconstructing the optics model with the likelihood function
  Eq.~\eqref{eq:likelihood}, and the prior distribution \eqref{eq:prior}
  and \eqref{eq:prior2}.  The predicted optics parameters with their
  uncertainties were then calculated based on another likelihood function
  between quadrupoles and the locations of undulators with
  Eq.~\eqref{eq:prediction}.

\begin{figure*}
  \centering \includegraphics[width=1.\textwidth]{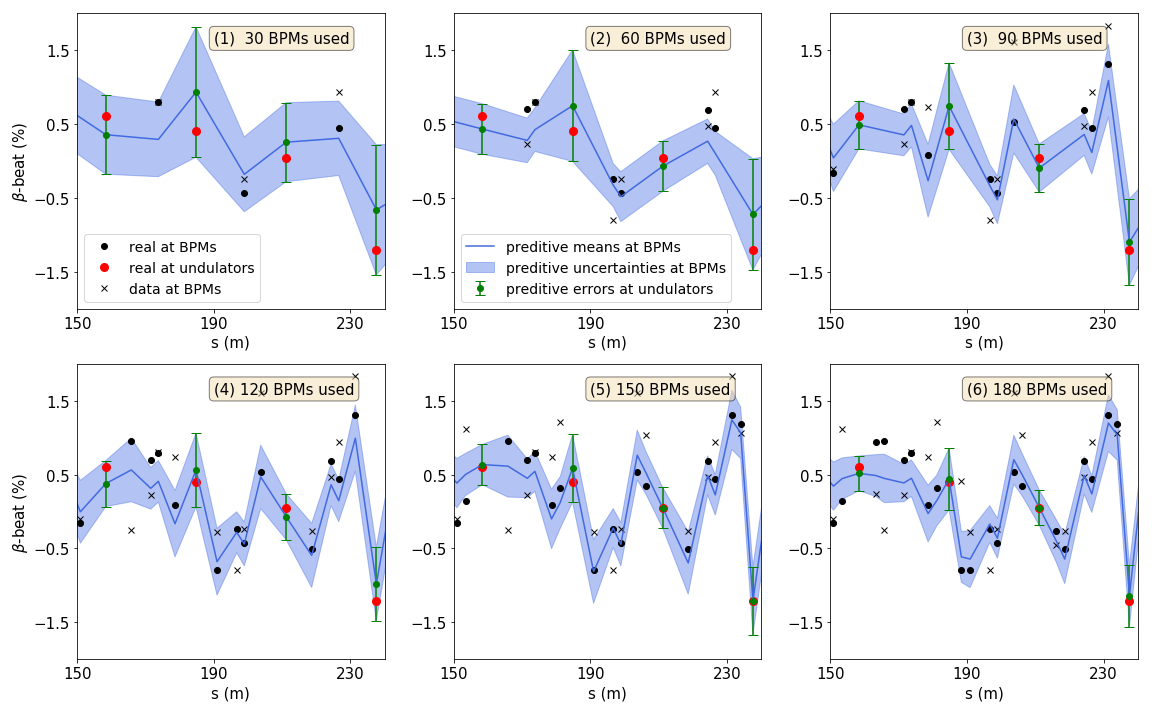}
  \caption{\label{fig:IDVsBPM} A zoomed-in view of predicted means and
    variances of $\beta$ observed at both BPMs (the training set) and
    undulators (POIs) for a section (spanning 3 cells, 4 POIs) of the
    NSLS-II ring. Black and red dots represent the real $\beta$ values at
    BPMs and POIs. Black crosses are the data observed by the BPMs.  The
    light blue lines with a shadow are the predictions at the BPMs, and
    the green error bars are the final prediction at POIs. From subplot 1
    to 6, the quantity of BPMs used increases gradually. A large set of
    training data (i.e. using more BPMs) for the regression does improve
    the accuracy and precision of the predicted results at POIs.  However,
    the improvement becomes less apparent after using more than 120 BPMs.}
  \end{figure*}

  The results of comparison are illustrated in Fig.\ref{fig:IDVsBPM}.  As
  with any regression problem, the training data distribution (i.e. the
  BPM locations) should be as uniform as possible within the continuous $s$
  domain.  There are 30 cells in the NSLS-II ring, and each cell has 6
  BPMs.  Equal numbers of BPMs were selected from each cell to make the
  training data uniformly distributed.  The goal was to predict all
  straight section optics simultaneously.  The predicted performance was
  therefore evaluated by averaging at multiple straight centers.
  Initially, one BPM was selected per cell.  The number of selected BPMs
  was then gradually increased to observe the evolution of predictive
  errors.  It was found that utilizing more BPMs improved the predicted
  performance, as expected.  Both the bias and uncertainty were reduced
  with the quantity of BPMs.  However, the improvement became less and
  less apparent once more than 4 BPMs per cell were used.  

  Since there are 6 BPMs per cell at the NSLS-II ring, we chose different
  BPM combinations. We found that some patterns/combinations of BPMs were
  better used to capture/measure these types of optics distortions. For
  example, each end of the straight sections needs one BPM to observe the
  ID, and at least one BPM needs to be located inside the achromat arc in
  order to observe the dipoles.  The distribution of the BPMs does not
  need to be uniform in the longitudinal $s$ direction, instead, they
  should be uniform along the betatron phase propagation.  Collider rings
  would see this effect more clearly due to the existence of interaction
  points. However, for most light source rings, including the NSLS-II
  ring, the phase propagation along the longitudinal direction is mostly
  quite linear in the longitudinal direction.

\begin{figure}[!ht]
  \centering \includegraphics[width=1.\columnwidth]{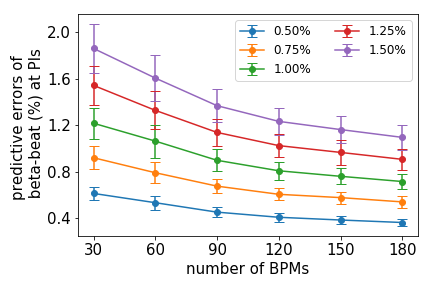}
  \caption{\label{fig:BPMprecision} Predictive $\beta$-beat errors
    (including bias and uncertainties) at the locations of undulator
    (POIs).  $\beta$s are observed with different number of BPMs and
    different resolutions. The resolution of $\beta$ is the final limit on
    predictive errors.  The higher the resolution each individual BPM has,
    the less number of BPMs are needed.}
  \end{figure}

  Next, we studied the effect of $\beta$ measurement resolution on the
  predictive errors.  A similar analysis was carried out but with
  different $\beta$-resolution as illustrated in
  Fig.~\ref{fig:BPMprecision}.  By observing Fig.~\ref{fig:BPMprecision},
  several conclusions can be drawn: (1) The degradation of the $\beta$
  resolution reduced the accuracy of the generalized optics model.
  However, this can be improved by applying a more complicated optics
  model~\cite{Hao:2019lmn}.  Thus, the BPM TbT resolution is the final
  limit on the resolution of $\beta$ parameters.  In order to accurately
  and precisely predict the beam properties at POIs, improving the
  resolution of BPMs is crucial.  (2) After a certain point, the predicted
  performance is not improved significantly with the quantity of BPMs as
  seen in both Fig~\ref{fig:IDVsBPM} and \ref{fig:BPMprecision}.  The
  advantage of reduction of predictive errors will gradually level out
  once enough BPMs are used.  Meaning that quantitatively, the improvement
  in error reduction will eventually become negligible compared to the
  cost of adding more BPMs.  The higher the resolution each individual BPM
  has, the less number of BPMs are needed.  There should be a compromise
  between the required quality and quantity of BPMs to achieve an expected
  predictive accuracy.  (3) The quality (resolution) is much more
  important than the quantity of BPMs from the point of view of optics
  characterization.  For example, at NSLS-II, in order to resolve 1\%
  brightness degradation, at least 120 BPMs with a $\beta$ resolution
  better than 1\% are needed, or 90 BPMs with a 0.75\% resolution, etc.
  Having more BPMs than is needed creates no obvious, significant
  improvement.  Having 60 high precision (0.5\% $\beta$-resolution) BPMs
  yields a better performance than having 180 low precision (1\%) BPMs in
  this example.

\section{\label{sect:summary} Discussion and summary}
  A systematic approach has been proposed to analyze a BPM systemís
  technical requirements in this manuscript.  The approach is based on the
  resolution requirements for monitoring a machine's ultimate performance.
  The Bayesian Gaussian regression is useful in statistical data
  modelling, such as reconstructing a ring's optics model from beam TbT
  data.  The optics properties of the ring are contained in a collection
  of data having a normal distribution.  From past experience in designing
  and commissioning various accelerators, many will intuitively realize
  that having more BPMs does not always significantly improve diagnostics
  performance and is therefore not necessarily cost-effective for an
  accelerator design. Using the Gaussian regression method, however,
  confirmed that quantitatively. More importantly, a reasonable compromise
  can be reached between the quality (resolution) and the quantity of BPMs
  using this method.

  It is worth noting that our approach is simplified as a linear
  regression by assuming a known linear dependence of optics distortion on
  quadrupole errors.  If a ring's optics are significantly different from
  the design model, this assumption is not valid. In our case, we needed
  to iteratively calculate the likelihood function $\bs{M}$ by
  incorporating the posterior mean of quadrupole errors
  Eq.~\eqref{eq:posteriorMean} and compare it to the optics model until
  the best convergence was reached.  This was not discussed in this paper,
  however, because our analysis applies best to machines whose optics are
  quite close to their design model.  Other important effects on X-ray
  brightness, such as quadrupolar errors from sextupole feed down, skew
  quadrupoles, longitudinal misalignments of quadrupoles and BPMs,
  systematic gain errors in BPMs, magnet fringe field etc. are not
  addressed in detail here.  These effects are neglected at the NSLS-II
  ring because either they are small compared with the quadrupole
  excitation errors and hysteresis, or their effects have been integrated
  into our optics model.  The Gaussian regression method outlined here,
  however, can be expanded to take them into account if necessary.

  In a ring-based accelerator, BPMs are used for multiple other purposes,
  such as orbit monitoring and optics characterization, etc.  In this
  paper we only concentrated on a particular use case of TbT data to
  characterize the linear optics, and then to predict X-ray beam
  brightness performance.  A similar analysis can be applied to the orbit
  stability, and dynamic aperture reduction due to $\beta$-beat as well.
  An accelerator's BPM system needs to satisfy several objectives
  simultaneously.  Therefore the Gaussian regression approach could/should
  be extended to a higher dimension parameter space to achieve an optimal
  compromise among these objectives.

\begin{acknowledgments}
  We would like to thank Dr. O. Chubar, Dr. A. He, Dr. D. Hidas and
  Dr. T. Shaftan (BNL) for discussing the undulator brightness evaluation,
  and Dr. X. Huang (SLAC) for other fruitful discussions.  This research
  used resources of the National Synchrotron Light Source II, a
  U.S. Department of Energy (DOE) Office of Science User Facility operated
  for the DOE Office of Science by Brookhaven National Laboratory under
  Contract No. DE-SC0012704. This work is also supported by the National
  Science Foundation under Cooperative Agreement PHY-1102511, the State of
  Michigan and Michigan State University.

\end{acknowledgments}
 
\bibliography{diag.bib}

\end{document}